\documentclass[12pt]{article}

\usepackage{epsfig}

\usepackage{amssymb}
\usepackage{amsfonts}

\def\be{\begin{equation}}
\def\ee{\end{equation}}
\def\ba{\begin{array}{c}}
\def\ea{\end{array}}

\newcommand{\kt}{\rangle}
\newcommand{\br}{\langle}

\newtheorem{thm}{Theorem}
\newtheorem{cor}[thm]{Corollary}
\newtheorem{lemma}[thm]{Lemma}

\begin{document}

\begin{center}

{\Large

Anomalous mechanisms of the loss of observability in non-Hermitian
quantum models

}

\vspace{0.8cm}

  {\bf Miloslav Znojil}

\vspace{0.2cm}

\vspace{1mm} The Czech Academy of Sciences, Nuclear Physics
Institute, Hlavn\'{\i} 130, 25068 \v{R}e\v{z}, Czech Republic

\vspace{0.2cm}

 and

\vspace{0.2cm}

Department of Physics, Faculty of Science, University of Hradec
Kr\'{a}lov\'{e},

Rokitansk\'{e}ho 62, 50003 Hradec Kr\'{a}lov\'{e},
 Czech Republic

\vspace{0.2cm}

 and

\vspace{0.2cm}

Institute of System Science, Durban University of Technology, P. O.
Box 1334, Durban 4000, South Africa;

\vspace{0.2cm}

 {e-mail: znojil@ujf.cas.cz}

\vspace{0.3cm}

and

\vspace{0.3cm}

{\bf Denis I. Borisov}

\vspace{0.2cm}

Institute of Mathematics CS USC RAS, Chernyshevskii str. 112, 450008
Ufa, Russia

\vspace{0.2cm}

 and

\vspace{0.2cm}

Bashkir State Pedagogical University named after M. Akhmulla,
October rev. str 3a, 450000 Ufa, Russia

\vspace{0.2cm}

 and

\vspace{0.2cm}

Department of Physics, Faculty of Science, University of Hradec
Kr\'{a}lov\'{e},

Rokitansk\'{e}ho 62, 50003 Hradec Kr\'{a}lov\'{e},
 Czech Republic

\vspace{0.2cm}

 e-mail: BorisovDI@yandex.ru

\end{center}

\newpage

\section*{Abstract}

Via several toy-model quantum Hamiltonians $H(\lambda)$ of a
non-tridiagonal low-dimensional
matrix form the existence of unusual observability
horizons is revealed. At the corresponding limiting values of
parameter $\lambda=\lambda^{(critical)}$ these new types of
quantum phase transitions
are interpreted as the points of confluence of several
decoupled Kato's exceptional points of equal or different orders.
Such a phenomenon of degeneracy of non-Hermitian
degeneracies seems to
ask for a reclassification of the possible
topologies of the complex energy
Riemann surfaces in the vicinity of branch points.

\subsection*{Keywords}
.

quantum phase transitions;

non-Hermitian Hamiltonians;

loss-of-observability mechanisms;

exceptional-point horizons;

degenerate exceptional points;

 \newpage

\section{Introduction}

The questions of stability and instability
belong
to the most fundamental features of reality solved and
resolved by quantum theory \cite{Messiah}.
From historical perspective
the formalism succeeded in
establishing a paradigmatic correspondence
between the
norm-preserving evolution
and the observability of the energy-operators {\it alias\,} Hamiltonians.
In the language of mathematics
one may recall the so called Stone theorem \cite{Stone} which connects
the unitarity of evolution with the
self-adjointness property  of Hamiltonians
$\mathfrak{h} =\mathfrak{h}^\dagger$
in a preselected
Hilbert space of states ${\cal L}$.

In 1998 Bender with Boettcher \cite{BB} turned the attention of
physics community to a half-forgotten possibility \cite{BG}
of having the {\em unitary\,} quantum evolution controlled,
in a less conventional Hilbert space (say, ${\cal K}$),
by a less conventional Schr\"{o}dinger equation
 \be
 {\rm i}\frac{d}{dt} \,|\psi(t)\kt = H\,|\psi(t)\kt\,,
 \ \ \ \ \ \ \ |\psi(t)\kt \in {\cal K}\,
 \ee
in which the Hamiltonian operator itself is {\em
manifestly non-Hermitian}, $H \neq H^\dagger$.
The related possibility of
circumventing the apparent limitations
attributed to the
Stone theorem gave birth to a powerful theoretical paradigm.
In its older applications in
condensed matter physics \cite{Dyson}
and in
nuclear physics \cite{Geyer}
the underlying, hiddenly Hermitian Hamiltonians
were called quasi-Hermitian (cf. also \cite{Dieudonne}).
Nowadays, the formalism is better known under the names of
${\cal PT}-$symmetric \cite{Carl} {\it alias\,}
pseudo-Hermitian \cite{ali} {\it alias\,}
three-Hilbert-space (3HS, \cite{SIGMA,NIP})
reformulation of quantum mechanics.

One of the main practical consequences of the
application of the upgraded paradigm to the concrete physical systems
(cf., e.g., the most recent summaries of the state of art in
\cite{book,Christodoulides,Carlbook})
may be seen in the fact that the change of the language
opened the way to a new analysis of the mechanisms of
the loss of the
quantum
stability and observability.
In essence, in contrast to the
Hermitian theory (in which the stability is ``robust'' because it
follows from the ``obligatory''
assumption that the Hamiltonian is, or
must be, self-adjoint), the work with non-Hermitian Hamiltonians
appeared open to {\em both\,}
of the ``robust'' and ``fragile'' possibilities.
Indeed, for the non-Hermitian Hamiltonians
which depend on some real parameters,
$H=H(a,b,\ldots)$,
the quantum system in question is, typically,
found stable for parameters lying
{\em strictly\,} inside
a ``physical'' domain of parameters ${\cal D}$
with a nontrivial boundary  $\partial {\cal D}$
(see, e.g., \cite{DDTsusy} for illustration).

One of the most impressive real-system
illustrations of the
extended descriptive capacity of
the 3HS formulation of quantum theory
may be found in papers \cite{[37],[38],Uwe,zaUwem}.
In these papers the authors generalized,
to a non-Hermitian but ${\cal PT}-$symmetric
(i.e., stable) version,
the conventional
Hubbard's \cite{Lieb} multi-bosonic
\cite{BHold} Hermitian model. Whereas
the traditional Hermitian version
of the model only became truly popular
after a discovery of its support of sophisticated,
difficult to localize \cite{cinaj}
phase transitions between
superfluid and Mott-insulator quantum phases
\cite{Mathew,[3],[4]},
the innovative non-Hermitian but
${\cal PT}-$symmetric Bose-Hubbard (BH) model
opened, {\it inter alia}, the possibility
of reaching the
very boundary
$\partial {\cal D}$ of the
physical domain of its parameters.
Even near this loss-of-observability boundary
the evolution remains unitary in a
way shown in
\cite{jinsong,corridors}.

Naturally, the
observability of the system
(i.e., the reality of the energies) becomes lost
when the parameters happen to cross the boundary
$\partial {\cal D}$.
In the special cases exhibiting the
parity-time symmetry (${\cal PT}-$symmetry)
one speaks about the spontaneous breakdown of this
symmetry, of particular interest in
quantum field theory \cite{BM}.
In general, even the mere
closeness of these boundaries
(called, in mathematics, exceptional points, EPs \cite{Kato})
opens, immediately, a lot of interesting new physics
(cf., e.g., \cite{locali}).
The theoretical scope and experimental realizations
of these phenomena
range from the
quantum Bose-Einstein condensation \cite{Uwe,Carta}
and from the action of spin-orbit interaction in coupled resonators
inducing exceptional points of arbitrary order
\cite{arbiorder}
up to some unusual effects in
classical acoustics reflecting,
in particular, the new mathematics
of the
emergence and coalescence of exceptional points
occurring in multiplets
\cite{realiza}.

In this context, our present study was motivated
by a fairly puzzling empirical observation that
virtually all of the phenomenologically meaningful
EP-related loss-of-observability boundaries
$\partial {\cal D}$ seem to belong to
the same, specific
subcategory of EPs at which the corresponding
(by definition \cite{Kato}, non-diagonalizable) limits
$H(a^{(EP)},b^{(EP)},\ldots)$ of the Hamiltonians
represent, in the words of one of the
typical physics-oriented studies,
``a peculiar type of non-Hermitian
degeneracy where a macroscopic fraction of the states
coalesce at a single point with a geometrical
multiplicity of one'' \cite{locali}.
Equivalently, this means that
matrices
$H(a^{(EP)},b^{(EP)},\ldots)$
prove mathematically user-friendly because,
using the words of Ref.~\cite{Uwe}, they
``are similar to a {\em single\,} ${N}$ by ${N}$ Jordan block''
 \be
 %{\cal J}^{({N})} \equiv
 J^{({N})}(\eta)=\left [\begin {array}{ccccc}
    \eta&1&0&\ldots&0
 \\{}0&\eta&1&\ddots&\vdots
 \\{}0&0&\eta&\ddots&0
 \\{}\vdots&\ddots&\ddots&\ddots&1
 \\{}0&\ldots&0&0&\eta
 \end {array}\right ] \,.
 %,
% \ \ \ \ {N} \geq 2\,
 \label{hisset}
 \ee
%with eigenvalue
%$\eta=0$''.
%
In the latter text
(which precedes equation Nr. 32 in {\it loc. cit.})
we emphasized here
the word  {\em single\,}
because we did not find, in the literature, {\it any\,}
sufficiently realistic illustrative
example of {\it any\,} less specific, generic
EP degeneracy characterized by the
geometric
multiplicity greater than one.
Equivalently, we did not find any illustrative model in which
the Hamiltonian matrices
$H(a^{(EP)},b^{(EP)},\ldots)$
would {\em not\,} be similar, at the loss-of-observability boundary
$\partial {\cal D}$, to a {\em single\,} ${N}$ by ${N}$ Jordan block
(\ref{hisset}).

In such a context we originally sought for any nontrivial EP loss-of-observability
model in which one would not end up with
the extreme geometric multiplicity
equal to one. In parallel, we felt
motivated by the suspicion that
there may exist a hidden mathematical correspondence
between the extreme multiplicity-one degeneracy
and the most manipulation-friendly \cite{Wilkinson}
tridiagonality of all of the known EP-supporting
Hamiltonians, with the above-mentioned BH model
being the prominent, though not the only one \cite{maximal}
exactly solvable example.

This led us to the formulation of the project.
We sought for ${N}$ by ${N}$ Hamiltonian-operator
matrices $H^{({N})}$ for which the EP
loss-of-observability
mechanism would be ``anomalous'', i.e.,
characterized by the geometric multiplicity greater than one.
As long as we were aware that
such a project may, presumably, require the study of
non-tridiagonal
matrices, we restricted our attention just to the
Hamiltonians which remain tractable non-numerically.
Thus, being also aware of the fact that
in the experimental setting the matrix dimensions ${N}$
need not be to large, we decided to direct our
attention to the tractable but already nontrivial
family of models with ${N}=6$.

Under this
constraint we will be able to complement the ``common'' choice of the
non-degenerate
Jordan block limit (\ref{hisset}) by its two alternatives, viz.,
by
the degenerate Jordan-block direct sums
 \be
 %{\cal J}^{(6)}=
 J^{(4+2)}({\eta})=\left[ \begin {array}{cccccc}
 {\eta}&1&0&0&0&0
 \\\noalign{\medskip}0&{\eta}&1&0&0&0
 \\\noalign{\medskip}0&0&{\eta}&1&0&0
 \\\noalign{\medskip}0&0&0&{\eta}&0&0
 \\\noalign{\medskip}0&0&0&0&{\eta}&1
 \\\noalign{\medskip}0&0&0&0&0&{\eta}
 \end {array}
 \right]\,.
 \label{[9]}
 \ee
and
 \be
 J^{(2+2+2)}({\eta})=\left[ \begin {array}{cccccc}
 {\eta}&1&0&0&0&0
 \\\noalign{\medskip}0&{\eta}&0&0&0&0
 \\\noalign{\medskip}0&0&{\eta}&1&0&0
 \\\noalign{\medskip}0&0&0&{\eta}&0&0
 \\\noalign{\medskip}0&0&0&0&{\eta}&1
 \\\noalign{\medskip}0&0&0&0&0&{\eta}
 \end {array}
 \right]\,
 \label{[9v]}
 \ee
corresponding to the
geometric multiplicities equal to two and three, respectively.

\section{EP-related mechanisms of the loss of observability}

%{A.2. The specific non-degenerate EPN-mediated losses of
%observability}

%\subsection{Tridiagonal BH model with geometric multiplicity one}

As we already emphasized,
one of the best
realistic simulations of the
process of the loss of observability
is provided by the
three-parametric ${N}$ by ${N}$ matrix
${\cal PT}-$symmetric Bose-Hubbard Hamiltonian
$H^{({N})}(\gamma,v,c)$
of Ref.~\cite{Uwe}.
Indeed,
after an arbitrary choice of the dimension ${N}$
(related to the
number of bosons $\hat{N}={N}-1$ in the quantum system in question)
this Hamiltonian
offers a non-numerical, exactly solvable
toy model exhibiting
the non-Hermitian ${N}-$tuple EP-conditioned
degeneracy of the whole ${N}-$plet of the real-energy
observable eigenvalues
with the geometric
multiplicity equal to one.
For this class of models the
EP-related mechanisms of the loss of observability
is already known (see, e.g., Refs.~\cite{corridors,entropy}
and a brief description of their results
in Appendix A below).
It seems also worth adding that
all of the similar
tridiagonal-matrix
models admit also a
straightforward and recurrently constructive \cite{dieudeqsol}
unitary-evolution probabilistic
interpretation (for a concise outline
of the corresponding
version of
abstract quantum theory see Appendix~B below).

\subsection{Conventional, {\em non-degenerate-EP\,} structure}

For parameter-dependent Hamiltonians $H^{(N)}=H^{(N)}(\lambda)$, any
set of eigenvectors
(cf. Eq.~(\ref{ep}) in Appendix A below)
ceases to form a complete basis at
$\lambda=\lambda^{(EP)}$. In such a limit the degeneracy of
eigenvalues is accompanied by the degeneracy of eigenvectors
\cite{Kato}. In the non-degenerate, maximally non-diagonalizable
extreme
with $\lambda \to \lambda^{(EP)}= \lambda^{(EPN)}$ one has
 \be
 \lim_{\lambda \to \lambda^{(EPN)}}E_n(\lambda) = \eta\,,\ \ \ \ \
 \lim_{\lambda \to \lambda^{(EPN)}}|\psi_n(\lambda)\kt = |\Phi\kt\,,
  \ \ \ \ \
 n=0, 1, \ldots, N-1\,
 \label{degeene}
 \ee
(see, e.g.,
\cite{maximal}) so that the conventional time-independent bound-state
Schr\"{o}dinger equation $H^{(N)}(\lambda)\, |\psi\kt = E\,|\psi\kt$
ceases to be solvable. It can
only be replaced by an alternative relation
 \be
 H^{(N)}(\lambda^{(EP)})\, Q
 = Q\,{\cal J}^{(N)}(\eta)\,
 \label{GCrealt}
 \ee
with
the so called transition matrix $Q$
and with
a suitable, purpose-dependent non-diagonal, one-parametric
matrix ${\cal J}^{(N)}(\eta)$ as sampled here by Eqs.~(\ref{hisset}),
(\ref{[9]}) and (\ref{[9v]}).

In different contexts and applications, various
versions of the latter, ``canonical representation'' matrix can be
found in the literature (cf., e.g., \cite{Erkki}).
Most often they are being chosen in the form of Jordan block (\ref{hisset}).
Such a matrix remains non-diagonalizable so that it
cannot be treated as a
quantum observable anymore. It can only play the role of an
EP-related substitute for the spectrum.
In parallel, the matrices of associated
solutions $Q$ are called transition matrices. They may be perceived
as a formal EP analogue of the set of conventional wave functions.

\subsection{Unconventional, {\em degenerate-EP\,} structure EP2+EP2+EP2}

%\subsection{The phenomenon of confluence of exceptional points}

For the sake of definiteness let us consider the
``next-to-trivial'' Hilbert space with dimension $N=6$. Let us also
assume that our matrix toy-model Hamiltonian is real, asymmetric and
non-tridiagonal. Moreover, in Eq.~(\ref{GCrealt})
the elementary Jordan block of
Eq.~(\ref{hisset}) will be excluded as ``the known case''. Under
these specifications the usual EP-related degeneracy of the bound
state energies
  \be
  \lim_{\lambda \to \lambda^{(EP)}}E_n(\lambda) = \eta\,,
  \ \ \ \ \
 n=0, 1, 2,3,4, 5\,
 \label{fufu}
  \ee
may be accompanied by the following nontrivial decoupling of the EP
limits of the wave-function solutions of our Schr\"{o}dinger
Eq.~(\ref{ep}),
 \be
 %\lim_{\lambda \to \lambda^{(EP)}}E_n(\lambda) = \eta\,, \ \ \ \ \
% n=0, 1, 2,3,4, 5\,,
% \ \ \ \ \
 \lim_{\lambda \to \lambda^{(EP)}}|\psi_{k_j}(\lambda)
 \kt = |\Phi_j\kt\,,
 \ \ \ \ \
 k_j=2j-2,\,2j-1\,,
 \ \ \ \
 j=1,2,3\,.
 \label{degeef}
 \ee
Thus, the ``known'', EP6-related loss-of-observability pattern
(\ref{degeene}) becomes replaced by its anomalous EP2+EP2+EP2
alternative.

Formally, the upgrade may be characterized by the
replacement of Eq.~(\ref{hisset}) by another elementary option (\ref{[9v]}).
In this matrix (mimicking a triply degenerate exceptional point at
the observability horizon) the auxiliary parameter ${\eta}$
(i.e., the value of the energy in Eq.~(\ref{fufu})) may be
arbitrary.

\section{Elementary models with three degenerate EP2s\label{se3}}

\subsection{Toy model EP Hamiltonian with two diagonals}

Our first illustrative non-tridiagonal toy model
has the most
elementary cross-diagonal form
 \be
 H^{(2+2+2)}({\varepsilon})=
\left[ \begin {array}{cccccc} {\varepsilon}-5&0&0&0&0&5
\\\noalign{\medskip}0&{\varepsilon}-3&0&0&-3&0
\\\noalign{\medskip}0&0&{\varepsilon}-1&1&0&0
\\\noalign{\medskip}0&0&-1&{\varepsilon}+1&0&0
\\\noalign{\medskip}0&3&0&0&{\varepsilon}+3&0
\\\noalign{\medskip}-5&0&0&0&0&{\varepsilon}+5
\end {array} \right]\,.
 \label{[15]}
 \ee
Having chosen ${\varepsilon}=0$ (this value
represents, after all, just an inessential shift of the scale)
the insertion in Eq.~(\ref{GCrealt})
yielded
the solution with transition matrix
 $$
 Q=
 \left[ \begin {array}{cccccc} -5&1&0&0&0&0\\\noalign{\medskip}-3&0&-3
&0&-3&0\\\noalign{\medskip}-1&1&-1&1&0&0\\\noalign{\medskip}-1&0&-1&0&0
&0\\\noalign{\medskip}3&1&3&1&3&1\\\noalign{\medskip}-5&0&0&0&0&0
\end {array} \right]~\,
 $$
and with
the block-diagonal,
three times degenerate EP2+EP2+EP2 structure of the Jordan
matrix of Eq.~(\ref{[9v]}) with $\eta=\varepsilon=0$.
This confirmed our expectations
that the search for non-numerical six by six real matrix
models with degenerate EPs
should employ the non-tridiagonal candidates for the Hamiltonians.

\subsection{A tentative inclusion  of perturbations}

Elementary structure of the
anomalous EP-limit
matrix (\ref{[15]})
inspires a move towards
its three-parametric perturbed partner
%
%nejak to zkuplujme!
%
%% A6 :=
%>    <<-5,0,0,a,0,-5*b>|
%>    <0,-3,-a,0,3*b,0>|
%>    <0,a,-1,-b,0,a>|
%>    <-a,0,b,1,-a,0>|
%>    <0,-3*b,0,a,3,0>|
%>    <5*b,0,-a,0,0,5>
%>    >;
%> latex(%);
%>  h6:=Eigenvalues(A6);
 \be
  H^{(2+2+2)}(a,b,c)
  =\left[ \begin {array}{cccccc} {{}}-5&0&0&-a&0&c
 \\\noalign{\medskip}0&{{}}-3&a&0&-3\,b&0
 \\\noalign{\medskip}0&-a&{{}}-1&b&0&-a\\\noalign{\medskip}a&0
 &-b&{{}}+1&a&0\\\noalign{\medskip}0&3\,b&0&-a&{{}}+3&0
 \\\noalign{\medskip}-c&0
 &a&0&0&{{}}+5
 \end {array} \right]\,.
 \label{hapr}
 \ee
%h6m[4]:=subs(a=.3,h6[4]);

%plot({Re(h6m[1]),Re(h6m[2]),Re(h6m[3]),Re(h6m[4]),
%Re(h6m[5]),Re(h6m[6])},b=-1.3..1.3,
%>
%> axes=framed,color=black);

%Cardano formule (viz Figs. \ref{lo6ja3} a  \ref{lo6ja3b})

\begin{figure}[h]                    %instead of \begin{figure}[t]
\begin{center}                         %instead of \begin{center}
\epsfig{file=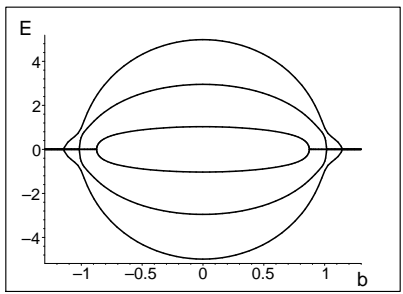,angle=270,width=0.30\textwidth}
\end{center}    % \sidecaption
%instead of \end{center}
\vspace{2mm} \caption{Real parts of the eigenenergies of
Hamiltonian $ H^{(2+2+2)}({0},0.3,b,5b)$ of Eq.~(\ref{hapr}).
 \label{lo6ja3}
 }
\end{figure}

 \noindent
The bound-state spectrum can be
obtained non-numerically, in the well known closed form called
Cardano formulae.
The existence of such a solution facilitates our
analysis, indeed. At the same time, the fully general results of
such a type already become long,
non-transparent and
unsuitable for an explicit printed
display. Graphical samples offer a
more appropriate form of presentation of the
key features of the dependence of the
energy levels $E_n$,
real or complex, on our three variable parameters.
Thus, the graphical sample of the spectrum as provided by
Fig.~\ref{lo6ja3} offers a nice example
of the unfolding of the
anomalous EP degeneracies of the preceding subsection.
At a
not too small fixed value of $a=3/10$, the six
real energy levels remain real in a fairly large interval of $b$
in the regime where $c=5\,b$. The
picture still keeps
traces of the unfolding of the degeneracy of the energies
near the two EP2+EP2+EP2 singularities which are localized at
the critical values of $b=\pm 1$. Far from these singularities the
spectrum looks robust and not too sensitive to the perturbations
controlled by the second-diagonal couplings $b$ and $c$.

\begin{figure}[h]                    %instead of \begin{figure}[t]
\begin{center}                         %instead of \begin{center}
\epsfig{file=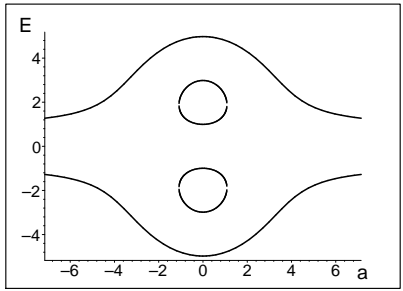,angle=270,width=0.30\textwidth}
\end{center}    % \sidecaption                      %instead of \end{center}
\vspace{2mm} \caption{Real spectrum of Hamiltonian $
H^{(2+2+2)}(a,b,5b)$ of Eq.~(\ref{hapr}) at a fixed value of
$b=1/10$.
 \label{lo6ja3b}
 }
\end{figure}

Figure \ref{lo6ja3b} complements the latter
visualization of the spectrum by a cross section
through the energy surface at $b=1/10$,
i.e., far from the critical value. This picture makes it clear that
even in the regular dynamical regime the stability of the
inner four excited states may be fragile. In particular,
the stability appears
sensitive to larger perturbations controlled by parameter $a$.

\subsection{Another, modified model of the loss of observability}

The more or less accidental choice of the parameter-dependence
of the perturbations should be complemented by the study of its
alternatives. Most obviously, the $a-$dependence of the energies could
be modified and/or combined with an independent variation of $c$. Thus,
we decided to analyze a modified, tilded form of
Hamiltonian
 \be
 \widetilde{H^{(2+2+2)}}(a,b)=\left[ \begin {array}{cccccc}
  -5&0&0&-5\,a&0&5\,b
 \\\noalign{\medskip}0&-3&3\,
 a&0&-3\,b&0\\\noalign{\medskip}0&-3\,a&-1&b&0&-5\,a
\\\noalign{\medskip}5\,a&0&-b&1&3\,a&0\\\noalign{\medskip}0&3\,b&0&-3
\,a&3&0\\\noalign{\medskip}-5\,b&0&5\,a&0&0&5\end {array} \right]\,.
 \label{rehapr}
 \ee
At a fixed $b$ the spectral locus appeared to be
entirely different, composed of a
vertical triplet of slightly deformed circles (presentation of such
an elementary picture would be redundant). At a variable $b$ the
real energies appeared to form three vertically arranged separate
tubes. They were found to merge, not too surprisingly, at the
EP2+EP2+EP2 ends with $b=b^{(EP)}_\pm=\pm 1$.

%Alternative confluences of exceptional points

\section{Models with the geometric multiplicity equal to two\label{se4}}

Obviously, our two illustrative examples demonstrated that
our trial and error search of degenerate EPs was successful when
based on the transition from tridiagonal to more-diagonal
sparse matrices of Hamiltonians. Now, let us apply the same
model-building strategy to some more complicated
matrix-element arrangements.

\subsection{The first two-block illustration: EP4 + EP2}

After we skipped the not too interesting EP3+EP3 scenario we decided to
search for the confluence EP4+EP2,
i.e., for a merger of the two exceptional
points of different orders using, this time, the direct sum
(\ref{[9]}) of two
Jordan-block submatrices.
With the most comfortable choice of shift ${\varepsilon}=0$
our search resulted in
the following
non-tridiagonal and manifestly non-Hermitian limiting
(i.e., unphysical, non-diagonalizable) EP4+EP2 Hamiltonian
 \be
 H^{(4+2)}=\left[ \begin {array}{cccccc}
  -9&3\,\sqrt {3}&0&0&0&0
 \\\noalign{\medskip}-3\,
 \sqrt {3}&-3&0&0&6&0
 \\\noalign{\medskip}0&0&-1&1&0&0
 \\\noalign{\medskip}0&0&-1&1&0&0
 \\\noalign{\medskip}0&-6&0&0&3&3\,\sqrt {3}
 \\\noalign{\medskip}0&0&0&0&-3\,\sqrt {3}&9
 \end {array} \right]\,.
 \label{filha}
 \ee
This Hamiltonian may be assigned the following well-behaved
transition matrix
  $$
  Q=\left[ \begin {array}{cccccc} -162&54&-9&1&0&0\\\noalign{\medskip}-162\,
  \sqrt {3}&36\,\sqrt {3}&-3\,\sqrt {3}&0&0&0\\\noalign{\medskip}0&0
&-1&1&-1&1\\\noalign{\medskip}0&0&-1&0&-1&0\\\noalign{\medskip}-162\,
\sqrt {3}&18\,\sqrt {3}&0&0&0&0\\\noalign{\medskip}-162&0&0&0&0&0
\end {array} \right]
  $$
with non-vanishing determinant $\det Q=26244$.
Again, our conclusion
is that
even the rather naive trial and error
search for the anomalous EP candidates may succeed.
Moreover, even in the more complicated EP4+EP2 arrangement
the search reveals the existence
of very sparse candidate matrices,
provided only that they are chosen more-than-tridiagonal.

What should follow would be
an
explicit construction of some perturbations compatible
with the reality of the spectrum, and describing the
path-of-unitarity trajectories
of the loss of observability.
Nevertheless,
the present elementary six by six matrix form
of our unperturbed Hamiltonian
would certainly
make these steps
an elementary, redundant and boring exercise
in linear algebra.

Naturally, the task would be
mathematically more difficult
at the higher matrix dimensions. Incidentally,
in the light of the well known related experimental-simulation
challenges \cite{Carta} there is no really
urgent need of any extensive and exhaustive generalization.
Just the display of a few characteristic
illustrative examples seems sufficient
for the inspiration of experimentalists
at present.

%\subsection{The second two-block illustration: EP4 + EP2}
\subsection{Another EP4 + EP2 model}

In the same spirit as
above let us now introduce the following two-parametric perturbed
Hamiltonian of a minimal-coupling type,
 \be
 H^{(6)}(\tau,\beta)=
  \left[ \begin {array}{cccccc} -9&3\,\sqrt
 {3-3\,{\tau}}&0&0&0&0\\\noalign{\medskip}-3\,\sqrt
 {3-3\,{\tau}}&-3&{\it \sqrt{\beta}}&0&-6\,\sqrt {1-{\tau}}&0
 \\\noalign{\medskip}0&-{\it \sqrt{\beta}}&-1&\sqrt {1-{\tau}}&0&0
 \\\noalign{\medskip}0
 &0&-\sqrt {1-{\tau}}&1&{\it
 \sqrt{\beta}}&0\\\noalign{\medskip}0&6\,\sqrt {1-{\tau}}&0&-{ \it
 \sqrt{\beta}}&3&3\,\sqrt
 {3-3\,{\tau}}\\\noalign{\medskip}0&0&0&0&-3\,\sqrt {3-3
 \,{\tau}}&9\end {array} \right] \,.
 \label{domham}
 \ee
In the decoupled case with $\beta=0$ the secular equation
  $E^6-91\,{\tau}\,E^4+819\,{\tau}^2\,E^2-729\,{\tau}^3=0$
becomes solvable exactly yielding the parabolic $\tau-$dependence of
the six real bound state energies
 $$
 E(\tau)=
  \pm \sqrt{\tau},\pm 3\,\sqrt{\tau},\pm 9\,\sqrt{\tau}\,.
  $$
They merge, as required, in the EP=EP4+EP2 limit of vanishing $\tau \to 0$.
In the general regular dynamical regime with $\beta \neq 0$, the
secular equation for energy $E=\pm \sqrt{{s}}$ is more complicated
but still cubic,

%
%tg:=solve(hyg,en):
%> plot3d({Im(tg[1]),Im(tg[2]),Im(tg[3]),
%Re(tg[1]),Re(tg[2]),Re(tg[3]),100000/{\beta}},
%> t=0..0.2,{\beta}=-0.2..0.15,view=0.001..0.0015,
%> axes=boxed,orientation=[-89.7,0.285],numpoints=10000,color=black);

\begin{figure}[h]                    %instead of \begin{figure}[t]
\begin{center}                         %instead of \begin{center}
\epsfig{file=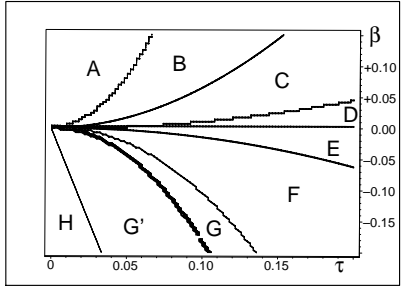,angle=270,width=0.30\textwidth}
\end{center}    % \sidecaption
%instead of \end{center}
\vspace{2mm} \caption{Numerically determined curves of zeros of the
real and/or imaginary parts of the roots $s_n$ of secular equation
(\ref{sekek}). An auxiliary line of $\beta=0$ has been added to guide
the eye,
and the curve separating the domains $G$ and $G'$ should
be ignored as
a spurious artifact of the numerics.
%
%Nuly 3 funkci ${{s}}$ (fazovy diagram near EP4+EP2)
 \label{6ja4b}
 }
\end{figure}

%%\newpage
% hyg:=charpoly(A6x,v);
%aa:=sqrt({\beta});v:=sqrt({{s}});
%tg:=solve(hyg,{{s}}):
%plot3d({Im(tg[1]),Im(tg[2]),Im(tg[3])},
%t=0..0.2,{\beta}=-0.2..0.15,view=0.01..0.015,
%axes=boxed,orientation=[-90.3,0.285],numpoints=5000);
%
%
%\begin{figure}[h]                    %instead of \begin{figure}[t]
%\begin{center}                         %instead of \begin{center}
%\epsfig{file=imag0.eps,angle=270,width=0.36\textwidth}
%\end{center}    % \sidecaption
%instead of \end{center}
%\vspace{2mm} \caption{Nuly 3 funkci Im E.
% \label{6ja3b}
% }
%\end{figure}

%plot3d({Re(tg[1]),Re(tg[2]),Re(tg[3])},
%t=0..0.2,{\beta}=-0.2..0.15,view=-0.01..0.015,
%axes=boxed,orientation=[-90.2,0.285],numpoints=5000);

%
%\begin{figure}[h]                    %instead of \begin{figure}[t]
%\begin{center}                         %instead of \begin{center}
%\epsfig{file=realy.eps,angle=270,width=0.36\textwidth}
%\end{center}    % \sidecaption
%instead of \end{center}
%\vspace{2mm} \caption{Nuly 3 funkci Re E.
% \label{6ja4b}
% }
%\end{figure}

%
%
%\begin{figure}[h]                    %instead of \begin{figure}[t]
%\begin{center}                         %instead of \begin{center}
%\epsfig{file=total.eps,angle=270,width=0.36\textwidth}
%\end{center}    % \sidecaption
%instead of \end{center}
%\vspace{2mm} \caption{Nuly 3 funkci E (fazovy diagram near EP4+EP3)
% \label{6ja4b}
% }
%\end{figure}

%$\hat{\lambda}(t)$
%\begin{widetext}
\begin{table}[h]
\caption{Conditions of unitarity of evolution for Hamiltonian
(\ref{domham}).
 }
\vspace{0.5cm}
 \label{owe}
\centering {\small
\begin{tabular}{||c||c|c||}
\hline \hline
  \multicolumn{1}{||c||}{\rm {\rm {\rm domain } }}
   &\multicolumn{2}{c||}{\rm {\rm {\rm condition (levels
   $n=0,1,2$):}
   }}
     \\
 %\hline
  \multicolumn{1}{||c||}{\rm {\rm {\rm  (see Fig.~\ref{6ja4b}) } }}
   &\multicolumn{1}{c}{{\rm  Im} ${{s}}_n(\beta,\tau) = 0$,}
    &\multicolumn{1}{c||}{{\rm  Re} ${{s}}_n(\beta,\tau) >
   0$}
     \\
 \hline
 \hline
  {\rm A} &{\rm  satisfied }&{\rm  not satisfied}
 \\
  {\rm B} &{\rm  not  satisfied }&{\rm  not satisfied}
 \\
  {\rm C} &{\rm  not  satisfied }&{\rm  satisfied}
 \\
 \hline
  {\rm D} &{\rm  satisfied }&{\rm  satisfied}
 \\
  {\rm E} &{\rm  satisfied }&{\rm  satisfied}
 \\
 \hline
  {\rm F} &{\rm  not  satisfied }&{\rm  not satisfied}
 \\
  {\rm G/%} &{\rm  not  satisfied }&{\rm  not satisfied}
% \\
%  {\rm
G'} &{\rm  not  satisfied }&{\rm  not satisfied}
 \\
  {\rm H} &{\rm  not  satisfied }&{\rm  not satisfied}
 \\
 \hline \hline
\end{tabular}}
\end{table}
%\end{widetext}

 \noindent
 \be
 {{\it {{s}}}}^{3}+ \left( -91\,{\tau}+2\,{\it {\beta}} \right)
  {{\it {{s}}}}^{2}+
 \left( -114\,{\it {\beta}}+819\,{{\tau}}^{2}-42\,
 {\tau}{\it {\beta}}+{{\it {\beta}}}^{2}
 \right) {\it {{s}}}-729\,{{\tau}}^{3}-486\,
 {\tau}{\it {\beta}}-81\,{{\it {\beta}}}^{2}=0\,.
 \label{sekek}
 \ee
This equation is Cardano-solvable so that it may be again expected
to determine all of the relevant properties of the energy spectrum,
in principle at least. In practice, however, the similar answers of
a complicated algebraic form remain as clumsy as their above-studied
predecessors. Fortunately, the remedy remains the same as before.
Our graphical description of the energy-level curves
$E_n(\tau,\beta)$ with $n=0,1,2,3,$ and $5$ may be made available as
fully two-parametric and, in this sense, exhaustive.

The price to pay is that our results must be now presented in a
combination of Figure \ref{6ja4b} with Table~\ref{owe}. Indeed,  the
Table is needed to endow the candidates for the energy curves with
the physical meaning as well as with the phenomenological
acceptability. One can conclude that the quantum system in question
remains unitary (i.e., stable {\it alias}, in the sense of
Ref.~\cite{Geyer}, quasi-Hermitian) only inside the subdomains D and
E of the whole plane of parameters. Only inside the ``corridor'' D +
E our secular equation (\ref{sekek}) has the full triplet of the
real and positive roots $s_n$ yielding the six acceptable (i.e.,
real) bound-state energies $E^{(\pm)}_n=\pm \sqrt{s_n}$.

Let us add that the picture contains a ``redundant'', $\beta=0$ line
which was inserted by hands. Guiding the eye, this line separates
the two $\beta-$sign-different but, otherwise, fully
unitarity-supporting access-to-EP subcorridors $D$ and $E$. We may
summarize:

\begin{itemize}

\item
the picture and table clearly demonstrate the existence and
uniqueness of a non-empty unitarity-compatible corridor $
{\cal D}=$ D + E of
quantum stability;

\item
this corridor ${\cal D}$  connects the interior of physical domain of parameters
with its EP extreme of maximal-non-Hermiticity;

\item
the boundary between D and E is artificial, marking the change of
sign of $\beta$ in parameter $\sqrt{\beta}$. In this sense the line
$\beta=0$ separates the non-Hermitian and Hermitian versions of the
Hamiltonian.

%
%\item
%the curve separating the domains G and G' is a purely formal
%artifact. Its presence reflects just an irrelevant round-off
%singularity in the Cardano formulae.

\end{itemize}

%
%One of the most interesting features of the spectrum of energies of
%our last illustrative example near its EP4+EP2 loss-of-obsservabilty
%extreme at $\beta=\tau=0$ ...

%\newpage

%The phenomenon of non-Hermitian degeneracy

%

%{A word of warning}

\subsection{Robust degeneracies and non-perturbative phenomena}

In our paper we reminded the readers that whenever we modify the
Hamiltonians, what follows is a change of the operator of the metric
defining the correct physical Hilbert space of states ${\cal H}$.
This means that the standard constructive recipes of perturbation
theory have to be modified \cite{entropy}. In the
single-Jordan-block models  a truly instructive illustration of some
slightly counterintuitive consequences may be found in
Refs.~\cite{Uwe} and \cite{admissible}. Also in the present,
``anomalous'' context of the degenerate-Jordan-block models, the
intuitive and/or hand-waving arguments may often fail as well.

In order to illustrate the danger, let us endow the above-mentioned
EP-limit model (\ref{filha}) with a particularly minimalistic
perturbation,
 \be
 H^{(4+2)} \to
 \widetilde{H^{(4+2)}}(\gamma)
 =\left[ \begin {array}{cccccc} -9&3\,\sqrt {3}&0&0&0&0
 \\\noalign{\medskip}-3\,
 \sqrt {3}&-3&0&0&6&0\\\noalign{\medskip}0&0&-1
 &1&0&0\\\noalign{\medskip}0&0&-1&1&0&0\\\noalign{\medskip}0&-6&0&-3\,{
 \it {\gamma}}&3&3\,\sqrt {3}\\\noalign{\medskip}0&0&0&0&-3\,\sqrt {3}&9
 \end {array} \right]\,.
 \label{refilha}
 \ee
To our great surprise such a perturbation does not remove the
degeneracy. It even changes, unexpectedly, the type of the EP
singularity. Indeed, our new matrix may be assigned the
perturbation-independent canonical Jordan-block representation
$J^{(6)}$ as well as the perturbation-dependent transition matrix
 $$
 Q= \left[ \begin {array}{cccccc} -486\,\sqrt {3}{\it {\gamma}}&54\,\sqrt {3}{
\it {\gamma}}&9\,\sqrt {3}{\it {\gamma}}&-3\,\sqrt {3}{\it
{\gamma}}&1/2\,\sqrt {3}{\it {\gamma}}&-1/18\,\sqrt {3}{\it
{\gamma}}\\\noalign{\medskip}-1458\,{\it {\gamma}}&0&45\, {\it
{\gamma}}&-6\,{\it {\gamma}}&1/2\,{\it
{\gamma}}&0\\\noalign{\medskip}0&0&0&0&-1&1
\\\noalign{\medskip}0&0&0&0&-1&0\\\noalign{\medskip}-1458\,{\it {\gamma}}&-
162\,{\it {\gamma}}&36\,{\it
{\gamma}}&0&0&0\\\noalign{\medskip}-486\,\sqrt {3}{ \it
{\gamma}}&-108\,\sqrt {3}{\it {\gamma}}&0&0&0&0\end {array} \right]
 $$
with determinant $\det Q=19131876\,{\gamma}^4$. This determinant is
$\gamma-$dependent and non-vanishing at any non-vanishing $\gamma
\neq 0$. At the same time, the transition matrix ceases to be
invertible (i.e., applicable) in the zero-perturbation limit $\gamma
\to 0$. In this limit the system keeps staying on the
loss-of-observability horizon but this horizon encounters an abrupt
change of its type, EP6 $\to$ EP4+EP2! Naturally (cf., e.g.,
Ref.~\cite{BH}), the detection of similar qualitative features of
non-diagonalizable matrices often lies beyond the capability of the
conventional numerical algorithms.

\section{Discussion}

The detailed constructive study of behavior of quantum systems near
the EP horizon is a purely numerical task in general. The dedicated
literature abounds with the specific phenomenological
implementations of the idea. They range from the abstract concepts
of the quantum phase transitions and quantum chaos (cf., e.g.,
review papers \cite{Heiss,Heissb}) up to many innovative experiments
in optics and photonic \cite{Miri} (note, e.g., that the light
can stop at an analogue of quantum horizon \cite{Nimrod}), or in
optomechanics \cite{optomech}, etc. Using some elementary toy models
one could even contemplate a purely schematic conceptual application
of the non-Hermitian Heisenberg picture near horizons in quantum
cosmology \cite{FabioBB318}.

The study of horizons remains basically numerical even if the Hamiltonian
itself is represented by an effective finite-dimensional $N$ by $N$
matrix $H^{(N)}(\lambda)$ with $N \geq 4$ \cite{Nje4}. The only
non-numerical exception admitting all finite matrix dimensions $N <
\infty$ seems to have emerged in the context of tridiagonal
Hamiltonian matrices (cf., e.g., \cite{Uwe,maximal} or
\cite{bh4a5}). In such a class of user-friendlier models the
horizons (occurring at a finite coupling constant
$\lambda=\lambda^{(critical)}$) have successfully been identified
with elementary Kato's \cite{Kato} exceptional points,
$\lambda^{(critical)}=\lambda^{(EP)}$.

Naturally, the deeply dynamical character of the latter
singularities (in review \cite{Heiss} they were called
``ubiquitous'') converted them recently in one of the central
subjects in theoretical efforts as well as in a variety of
experimental simulations. In the latter setting let us just recall,
{\it pars pro toto}, the fact that in the one-dimensional photonic
crystals one can construct an EP horizon at which ``two EPs \ldots
coalesce \ldots and create a singularity of higher order''
\cite{Katodale}.

In our present paper we found
all of the similar, EP-related
pieces of information truly inspiring.
Thus, we reopened the theoretical question of a
generalization of the theory in which the EP horizons would be
studied in more detail. Via a few examples we illustrated
that such an analysis seems truly rewarding,
especially when one
starts to study the more-than-tridiagonal matrices $H^{(N)}$.
By means of a detailed study of several toy-model matrices we
demonstrated that in  the more-than-tridiagonal cases
there emerges an
anomalous structure of the EPs. The Hamiltonians
cease to be represented by
the traditional, single canonical Jordan block
of maximal dimension $N$
but rather by a block-diagonal matrix with the number of
Jordan-block submatrices equal to the geometric multiplicity
of the EP singularity in question.

The existence of a wealth of these anomalous EP structures
has been revealed, tractable as a degenerate
superposition of several independent
exceptional points of lower orders.
Such a result might prove inspiring
for a continuing theoretical as well as experimental analysis.

\subsection{Experimental realizations of
access to EPs}

The formal mathematical construction of a
corridor ${\cal D}$ explains the mechanism of
the loss of the observability
and of the fall of a unitary
quantum system upon its EP singularity.
For a real-world applicability of the theory
these mechanisms must remain ``robust'',
i.e., reasonably insensitive to random perturbations.
A brief complementary comment seems necessary because
it is precisely this ``robustness''
assumption which was exposed to a harsh criticism
by mathematicians recently
(see, e.g., \cite{Trefethen,Viola}).

In essence, this conflict of opinions is just
an elementary
terminological
misunderstanding. The essence of the misunderstanding lies in the absence
of a clear specification of the underlying physics. In brief, one
could have said that the so called ``closed,''
unitary quantum systems remain ``robust'',
while the ``fragility''
exclusively applies just to the so called ``open'' quantum systems
living, in our present terminology, in the ``trivial-metric''
Hilbert space ${\cal K}$.
In this light the conflict is artificial: all of the
rigorous proofs and numerous illustrative
examples provided and discussed, e.g., in \cite{Trefethen,Viola}
are correct when applied to the
non-unitary, open quantum systems (cf. also our older
comment in
\cite{FR}).
Indeed,
the authors of the criticism worked just with a
simplified metric $\Theta = I$,
i.e., in ${\cal K}$.

From the perspective of the
closed, stable quantum systems the latter criticism
did not take into account the correct metric $\Theta \neq I$.
Far from EPs, in a weakly non-Hermitian dynamical regime
(as studied, e.g., in
\cite{weaknehe}) it would still make good sense to
analyze closed systems and replace, or
rather approximate, the correct but complicated physical Hilbert
space ${\cal H}$ by its manifestly unphysical but much
user-friendlier simplification ${\cal K}$. Then, indeed, a
``sufficient smallness'' of the perturbation specified via the
correct norm in ${\cal H}$ need not be too different from an
approximate specification via its norm in ${\cal K}$.

%{The requirement of the unitarity of evolution}

\subsection{Phenomenological applicability outlook}

Let us now return to the generic, ``non-pathological'' quantum
systems. The purpose of the analysis of the boundaries of their
observability and stability is, in general, twofold. Firstly, it can
lead to a basic intuitive understanding of the concept of an onset
of instabilities. Secondly, it can clarify the range of
applicability of mathematical methods. In this context, our present
introduction of several schematic models of physical reality was
motivated by both of these ambitions. In a concise comment on
applicability in physics we may add that among the quantum phenomena
explained via elementary models one may recall the phase transitions
in Bose-Hubbard systems \cite{BH} as well as the Big Bang models in
cosmology (for context see, e.g., an introductory report in
\cite{FabioBB318}, or section 5 in our recent review paper
\cite{LaraSymm323}).

In the methodical setting, in a direct continuation of our recent
studies \cite{FR,FMF,nje6}, just the elementary tools of linear
algebra may be expected to be needed again. Still, even these not
too complicated methods proved efficient enough to clarify that in
any sufficiently ambitious theoretical description of physical
reality an unexpectedly universal explanatory role proved played by
the Kato's notion of exceptional points.

Our present brief note opened several new EP-related directions of
possible future research. More specifically, we managed to suppress,
partially at least, the scepticism concerning the possibilities of
an efficient non-numerical tractability of multi-diagonal models as
expressed in paper \cite{prefertridi}. We revealed that in a way
guided by computer-assisted experimenting the non-numerical
descriptive capability of the more-than-tridiagonal Hamiltonians $H$
may be perceived more than promising.

Via our illustrative examples we demonstrated that in the future, a
transition to non-tridiagonal matrix models might prove also
productive from a purely phenomenological point of view. Rewarding,
as we explained, due to the capability of non-tridiagonal
Hamiltonian to mediate the mergers of individual EPs into
multiplets, with all of the not yet explored observable
consequences. Thus, the attention to the mergers could certainly
further enrich our current understanding of the ``unavoided
crossing'' phenomena \cite{ptho}, or of many other rather general
realizations of quantum phase transitions \cite{Denis}.

\section{Summary}

%\section{Anomalous horizons}

The model-building strategy based on the exclusive use of
tridiagonal matrices, Hermitian or not, can rely on several
advantages as sampled, e.g., in \cite{prefertridi}. Beyond the
tridiagonal models, due care must be paid to details. Even at the
not too large matrix dimensions $N$, one can sometimes be surprised
by unexpected mathematical difficulties \cite{Wilkinson}. One of the
obstacles appears particularly remarkable. It is characterized by a
rather rarely considered possibility of a deviation of our choice of
matrix ${\cal J}^{(N)}$ in Eq.~(\ref{GCrealt}) from its
conventional, frequently used single-Jordan-block version as
prescribed by Eq.~(\ref{hisset}). A few samples of such a deviation
were presented also in our paper.

In our considerations we felt influenced by the recent interest in
an experimental accessibility of a loss-of-observability
quantum-catastrophic processes. In fact, such an accessibility of the
EP-related horizon
of observability
is one of the most innovative and challenging consequences
of the innovative use of non-Hermitian operators of
observables with real spectra as studied in quasi-Hermitian
\cite{Geyer} {\it alias\,} PT-symmetric \cite{Carl} {\it alias\,}
crypto-Hermitian \cite{Smilga} {\it alias\,} three-Hilbert-space
\cite{SIGMA} {\it alias\,} pseudo-Hermitian \cite{ali}
representations of unitary quantum theory. In this context our present
detailed study of the structure of these horizons revealed that
the current preference of the
closed quantum systems with non-degenerate EPs
seems to be just an artifact which resulted from the
bigger constructive comfort provided by Hamiltonians
possessing a tridiagonal matrix representation.
Thus, a broad new field of research seems to be
open and waiting for a more systematic experimental as well
as theoretical analysis.

In our present paper the first steps were made in the spirit of
such a research project. Their constructive and computing part
involves the tests (and the empirical confirmations) of the
hypothesis of correspondence between the matrix tridiagonality
of the Hamiltonian and the non-degenerate nature of the EP
singularity. This confirmation was based on the use of the
non-tridiagonal but still exactly solvable EP2+EP2+EP2 examples
(\ref{[15]}), (\ref{hapr}) and (\ref{rehapr}) in section
\ref{se3}, and of their less elementary EP4+EP2 modifications
(\ref{filha}) and (\ref{domham}) in section \ref{se4}.

On a different, model-independent level of qualitative,
serendipitous discoveries we pointed out that whenever one
works with non-Hermitian matrices (which is, in the EP context
and applications, necessary), one must always carefully check
and test the possibilities of emergence of multiple not quite
expected technical subtleties. In this sense, a word of warning
was formulated and supported by an explicit ``ill-behaved''
illustrative matrix model (\ref{refilha}).

Last but not least, due to the not entirely standard overall theoretical
framework of our present paper, an enhancement of its clarity
in combination with a comparatively self-contained form has
been achieved by the inclusion of several review-like
explanatory texts relocated to the two Appendices.

\section*{Acknowledgements}

{Both authors acknowledge the financial support
by the Faculty of Science of the University of Hradec
Kr\'{a}lov\'{e}.}  MZ also acknowledges the support by the 
Excellence project P\v{r}F UHK 2212/2020.

\newpage

\section*{Appendix A. The mechanism of the loss of
observability in a small vicinity of a {\em non-degenerate\,} exceptional point}

During most studies of various versions of $N$ by $N$
matrices $H^{(N)}(\lambda)$ which are tridiagonal, the localization
of the Kato's exceptional points (of the maximal order
$N$, with $\lambda=\lambda^{(EPN)}$) was
achieved by non-numerical, purely algebraic methods
\cite{tridiagonal}. In the context of our
more ambitious and general present study, the
current state of art should be summarized briefly.

\subsection*{A.1. Tridiagonal matrix models}

 %The role of the concept of
%geometric multiplicity}

First of all, let us remind the readers that the construction of a
tridiagonal-matrix representation of a given realistic quantum
Hamiltonian operator is encountered, as an intermediate step, in the
majority of conventional numerical algorithms of solving the
bound-state Schr\"{o}dinger equations
 \be
 H^{(N)} \,|\psi_n\kt=E_n \,|\psi_n\kt
 \,,\ \ \ \ n = 0, 1, \ldots, N\,,
 \ \ \ \ N \leq \infty \,.
 \label{ep}
 \ee
Thus, in the context of numerical quantum mechanics the
tridiagonalization of matrices is a comparatively routine task. The
situation is different in the non-numerical versions of applied
linear algebra. In PT-symmetric quantum mechanics, in particular,
the matrix-tridiagonality requirement is often appreciated as
facilitating the constructive predictions of the results of
measurements~\cite{prefertridi}, mainly because these predictions
require the evaluation of mean values of the operators of
observables.

Under the assumption of the
non-Hermiticity of Hamiltonians $H^{(N)} \neq  \left [ H^{(N)}\right
]^\dagger$, tridiagonality is known to facilitate also  the
localization of the exceptional points forming, in the domains of
parameters, the phenomenologically deeply relevant
loss-of-the-observability boundaries \cite{maximal,Denis}.
Empirically, as we already mentioned, there seems to be a one-to-one correspondence between the tridiagonality of $H$
and the non-degenerate nature of its EPs characterized by the use of
the
Jordan blocks (\ref{hisset}) in Eq.~(\ref{GCrealt}).
For this reason
our present interest was concentrated on
non-tridiagonal matrix models with the
geometric multiplicities greater than one.

\subsection*{A.2. The losses of
observability due to small perturbations}

%The spontaneous breakdown of
%${\cal PT}-$symmetry}

Our recent
papers \cite{corridors,entropy} offered a detailed explanation of
the specific features of the processes of the loss of
observability
which,
in certain specific systems,
can be connected with the spontaneous breakdown of
supersymmetry \cite{DDTsusy}
or of
${\cal PT}-$symmetry \cite{Carlbook}.
Basically, these processes
correspond to the passage of the system through an EP
boundary $\partial {\cal D}$.
Naturally, once the system starts living in a small vicinity
of the EP singularity, it makes sense
to recall the formal tools of perturbation theory \cite{Kato}.

In the conventional, non-degenerate-EP-related dynamical regime
any Hamiltonian becomes close to its non-diagonalizable,
manifestly
unphysical EPN limit. Thus,
we have to consider the perturbed forms
 \be
 \widetilde{H}(\lambda)={J}^{(N)}(\eta)+g\,V^{(N)}(g)\,,
 \ \ \ \ \ g=g(\lambda)={\cal O}(\lambda -
\lambda^{(EP)})
 \label{refous}
 \ee
of its formal Jordan-block representations (\ref{hisset}).
The main conclusions
of the analysis were summarized in Ref~\cite{corridors}.
At an arbitrary matrix dimension $N$
the ``tilded'' model (\ref{refous})
is to be perceived as representing, under
certain conditions, an admissible Hamiltonian of a unitary quantum
system which remains stable under small perturbations.
In this language every process of the loss of observability
becomes realized inside a
corridor which connects the weakly non-Hermitian
dynamical regime
inside ${\cal D}$ with its EPN-related loss-of-observability boundary.

\begin{lemma}\cite{corridors}
\label{lej1} For the class of the real perturbation matrices
$V^{(N)}(g)$ with matrix elements which are uniformly bounded in
${\cal K}$, the eigenvalues of matrix $\widetilde{H}(\lambda)$ of
Eq.~(\ref{refous}) have the generic leading-order form
 \be
 E_j(\lambda) \ \sim \ {\rm a\ constant\,} +
 e^{2{\rm i}\pi j/N}\,\sqrt[N]{\lambda-\lambda^{(EPN)}}
  + \ldots\,,
 \ \ \ \ \ j = 0,1,  \ldots, N-1\,.
 \label{geeen}
 \ee
%which represents an immediate $N > 2$ generalization
%of Eq.~(\ref{ens2}).
\end{lemma}

\begin{cor}
For the generic perturbation of Lemma \ref{lej1} the spectrum of
energies can only be real at $N=2$.
\end{cor}

 \noindent
For $N>2$, the
specification of the class of the
admissible (i.e., ``sufficiently small'')
perturbations which would keep the system inside
the unitarity corridor ${\cal D}$ requires
further, rather severe
restrictions which may be found
described and proved in \cite{corridors}.

\begin{lemma}\cite{corridors}
For a guarantee of the reality of the spectrum of matrix
$\widetilde{H}(\lambda)$ of Eq.~(\ref{refous}) it is necessary to
guarantee that
 $V^{(N)}_{j+k,j}(g) ={\cal
O}(g^{(k-1)/2})\,$ at $k=1,2,\ldots,N-1\,$, at a sufficiently small
difference $g(\lambda)$, and at all $j$. \label{lej2}
\end{lemma}

\begin{cor}\cite{corridors,entropy}
Matrix $\widetilde{H}(\lambda)$ of Eq.~(\ref{refous}) may represent
a unitary, closed quantum system if and only if the spectrum is
real. At small $\lambda$ the acceptable parameters are then confined
to a comparatively narrow ``stability corridor'' forming an access
to the EPN extreme.
\end{cor}

\section*{Appendix B. The 3HS formalism {\it in nuce}}

%${\cal PCT}-$symmetry and its generalizations}

The equivalence between the conventional and
non-Hermitian-Hamiltonian descriptions of
unitary (i.e., stable) quantum systems  is based on
an {\it ad hoc\,} Hermitization of the operators of observables (cf. \cite{ali,SIGMA}). Let us now briefly outline the essence of the idea.

\subsection*{B.1. {\it Ad hoc\,} constructions of the physical inner products}

In a generic unitary quantum system let us assume that the
Hamiltonian is a finite-dimensional, $N$ by $N$ matrix with real
spectrum. Naturally,
it is necessary to preserve
a compatibility between the models with Hermitian
Hamiltonians
$\mathfrak{h}^{(N)}=\left(\mathfrak{h}^{(N)}\right)^\dagger$, and
those with non-Hermitian, quasi-Hermitian Hamiltonians $H^{(N)}$. In
the latter case it is in fact sufficient
to guarantee that there exists a positive definite and
Hermitian $N$ by $N$ matrix $\Theta=\Theta^\dagger$ such that
 \be
 \left (H^{(N)}\right )^\dagger\,\Theta=\Theta\, H^{(N)}\,
 \label{orequa}
 \ee
or, equivalently,
 \be
 H^{(N)}=\left (H^{(N)}\right )^\ddagger:=
 \Theta^{-1}\,
 \left (H^{(N)}\right )^\dagger\,\Theta
  %\equiv
 \,
 \label{orequb}
 \ee
(see, e.g., review \cite{Geyer}).

The discovery of merits of building quantum phenomenology using
quasi-Hermitian Hamiltonians is usually attributed to Dyson
\cite{Dyson} but it was only Bender with Boettcher \cite{BB} who
made the idea truly popular. They persuaded a broader physics
community about the fairly deep theoretical appeal of
quasi-Hermitian Hamiltonians (i.e., in their terminology, of the
non-Hermitian but ${\cal PT}-$symmetric Hamiltonians -- cf., e.g.,
detailed reviews \cite{Carl,book,Carlbook}).

In the language of physics
the concept of quasi-Hermiticity can be most easily interpreted
after a factorization of the matrix called metric,
 \be
 \Theta=\Omega^\dagger\Omega\,.
 \label{ore}
 \ee
This enables one to treat the quasi-Hermitan Hamiltonians as the
mere isospectral avatars of their standard textbook representations
 \be
 \mathfrak{h}=\Omega\,H\,\Omega^{-1}=\mathfrak{h}^\dagger\,.
 \label{orequ}
 \ee
Thus, the mapping (\ref{orequ}) may prove useful whenever the
conventional Hamiltonian $\mathfrak{h}$ (acting, say, in a
conventional Hilbert space ${\cal L}$) happens to be more
complicated than its partner $H$ acting in another Hilbert space
${\cal K}$.

In the theoretical physics of unitary quantum systems the
user-friendlier Hilbert space ${\cal K}$ plays an important
technical role. In spite of the obvious fact that the inner products
in ${\cal K}$ do not admit any immediate physical probabilistic
interpretation, an efficient remedy is straightforward. One simply
introduces the third, correct and physical Hilbert space ${\cal H}$
which only differs from ${\cal K}$ by an amended inner product,
 \be
 \br \psi_a|\psi_b\kt_{({\cal H})}=
 \br \psi_a|\Theta|\psi_b\kt_{({\cal K})}
 \,.
 \label{use}
 \ee
Thus, our (by assumption, positive definite and nontrivial) matrix
$\Theta \neq I$ is revealed to play the role of a non-trivial but
physics-determining inner-product metric.

\subsection*{B.2. The concept of smallness of
perturbations\label{nonvi}}

One of the most decisive advantages mediated by the choice of
quasi-Hermitian toy models may be seen in the fact that the
information about dynamics (encoded, in conventional models, into a
single Hermitian Hamiltonian operator or matrix $\mathfrak{h}$) is
now carried by the pair of operators (viz., by $H$ and $\Theta$). In
the literature explaining such a methodical advantage (cf., e.g.,
\cite{BB,Geyer}) people still often start the process of
model-building directly from a single non-Hermitian operator $H$
alone, especially when its real spectrum exhibits some
phenomenologically desirable descriptive properties (cf., e.g.,
\cite{Carl}).

The construction of partner operator $\Theta$ making the model
consistent is not always paid the same attention. Often, the duty is
only fulfilled formally, without recalling relation (\ref{orequa})
[or, equivalently, (\ref{orequb}) or (\ref{orequ})], and without
treating it as a constraint which has to restrict the freedom in our
choice of $\Theta$. In fact, the technical difficulty of the
construction of the metric redirected many physicists towards the
manifestly non-unitary, effective-model setups in which the
construction of the metric $\Theta(H)$ is not needed at all because,
in this setting, the role of the physical space is played, directly,
by the friendliest space, viz., by ${\cal K}$. Such a strategy is
also preferred, due to its simplicity, by mathematicians
\cite{Trefethen}.

In the alternative, technically more ambitious strategy aimed at the
quasi-Hermitian-operator description of the stable, unitarily
evolving quantum systems the key challenge lies in the proper
treatment of the subtle differences between the auxiliary and
physical Hilbert spaces ${\cal K}$ and ${\cal H}$, respectively
\cite{FR}. Such a challenge has two components. Firstly, one reveals
that the correct physical metric is necessarily
Hamiltonian-dependent. Even the physical Hilbert space itself must
be, therefore, expected to vary with the variation of the dynamics,
${\cal H}={\cal H}(H)$ \cite{FR}. Secondly, even at a fixed $H$ the
solution $\Theta=\Theta(H)$ of Eq.~(\ref{orequa}) remains ambiguous
\cite{lotor}.

The full strength of the challenge emerges during the studies of
stability, i.e., during the analysis of the influence of ``small''
perturbations
upon the processes leading, potentially,
to the loss of the observability of the system
\cite{entropy}. The problem has been clarified in
\cite{corridors} where we restricted our attention to the
Hamiltonians near their non-degenerate
EP singularity. Our detailed analysis of the
spectra of the tilded model (\ref{refous}) revealed that one cannot
define the acceptability of perturbations via a ``sufficient
smallness'' of the mere single coupling constant $g$. The reason
(discussed also
in preceding Appendix A) lies in a strongly hierarchically
anisotropic nature of {\em any available\,} metric $\Theta(H)$ at
{\em any\,} sufficiently small $g$. Such an anisotropy implies, in a
way illustrated in \cite{corridors}, that it is highly nontrivial to
evaluate the ``measurable size'' of the perturbation in
(\ref{refous}). In particular, this obstacle practically excludes a
guarantee of the smallness of the relevant norm of a perturbation in
${\cal H}$ using just its auxiliary, manifestly unphysical
representation in ${\cal K}$, i.e., working without a supplementary
reference to the structure of the metric \cite{entropy}.

\end{document}